# Observation of Fundamental Limit of Light Localization

Farbod Shafiei [1,*], Massoud R. Masir [1], Tommaso Orzali [2], Alexey Vert [,3], Man Hoi Wong [2], Gennadi Bersuker [4], Michael C. Downer [1]

[1] Department of Physics, The University of Texas at Austin, Austin TX 78712 USA
[2] SEMATECH, Albany NY 12203 USA
[3] Sunny Polytechnic Institute, Albany NY 12203 USA
[4] The Aerospace Corporation, Los Angeles CA 90245 USA
*farbod@physics.utexas.edu

**Abstract:**

**In disordered media light can be localized in the spaces between scattering sites which average to an optical mean free path (MFP). However the fundamental question of the smallest MFP that can support Anderson localization of light remains unanswered due to fabrication complexity of a scattering medium with controlled nano-scale gaps and lack of required resolution by far-field methods. Here we use scanning probe microscopy technique to collect localized light created at gaps between scattering crystallographic defects in a large variety set of nano-gap III-V medium. No localized spots correlated to MFP below ~14.5 nm is observed at second-harmonic collection at 390 nm. Experiment and simulation resulted in the first direct observation of suppression of Anderson light localization correlated to ~13 nm optical MFP that reveals a fundamental constraint in electromagnetism and photonics.**

## Introduction:

Scattering of light within disordered media can create intense, sub-wavelength concentrations of light due to interference and localization of the scattered fields. [John 1991, Wiersma 1997] The disorder leading to such optical localization can arise from crystallographic defects in semiconductors [Shafiei 2021], randomness in photonic crystals or media [Wiersma 2013, Sapienza 2010, Segev 2013, Wang 2020], size variations in micro/nanoparticle structures [Cao 1999, Sperling 2013, Schuurmans 1999, Mascheck 2012] or surface roughness in conductors. [Smolyaninov 1997]. Defects can play an important role in a variety of applications such as controlling polarization, increasing efficiency of heat-electricity conversion and pinning vortices in high temperature superconductors. [Hofling 2021, Kim 2015, Dam 1999] This raises the question of how to diagnose, understand and control such defects.

Understanding of electromagnetic waves at a deep sub-wavelength scale currently comes with significant challenges [Yang 2019, Rotenberg 2014]. Previous studies show that plasmonic enhancement of the light at metallic nano-junction is geometry dependent [Rotenberg 2014] and spatial limitation of electric fields can not be understood and observed as electric fields fill the nano-size gaps of the structures that are usually not directly accessible. [Savage 2012] In contrast, scattering and interference in Anderson localization in a disordered medium creates localization of electric fields even away from scattering sites (see details in 4th part of result section). Thus scattering and interference of light in Anderson localization provides an opportunity to better understand light nano-scale properties and spatial limitations if optical resolution and a disordered medium of these dimensions is available.

Localized light corresponding to an optical mean free path of a few hundred nanometers has previously been reported. [Schuurmans 1999, Lagendijk 2009, Lee 2018] This study addresses a still open question: What is the fundamental size limit of light localization that can be achieved within a disordered medium? While prevention of diffusion and localization of electron waves by disordered medium due to Anderson localization has been well studied [Anderson 1958, Lagendijk 2009], here we investigated a regime in which optical wave Anderson localization is suppressed by extreme space confinement. This question is relevant to optimization of super resolution microscopy [Mosk 2012, Willig 2006, Betzig 2006], unraveling the complex information buried in microscopy by optical deconvolution, image processing and pattern recognition. [de Aguiar 2017, Ritcher 2018, Stephens 2003, Feng 2019] While light scattering is a common technique to probe matter, suppression of light scattering not only it could provide insight to the foundation of electromagnetism, but could brought valuable information in the field such as ultra-cold atom. [ DeMarco 2021 ]

Here we report observations of light localization within 500 nm thick GaAs films that are grown on mis-matched Si substrates, and are thus pervaded by a high, but controllable, density of dislocation defects that act as optical nano-scale scattering sites.

Since this III-V medium has one of the strongest second-order nonlinear optical susceptibilities among naturally occurring materials, it selectively frequency-doubles the most intense concentrations of localized light. This natural selection feature greatly improves the ability of an uncoated fiber probe tip that collects second-harmonic light to discriminate intense concentrations of fundamental light, compared to a probe tip that collects light at the fundamental wavelength. [ Shafiei 2021, Shafiei 2022 ]   Consequently, test measurements of the resolution of our scanning fiber probe on nano-structured control samples demonstrated spatial resolution of ~19 nm when operating in second-harmonic collection mode (SD-1), collecting spots significantly smaller [Shafiei 2021] than the smallest localized intensity spikes reported previously. [ Lee 2018, Mascheck 2012 ] In contrast, diffuse scattered fundamental light dominated data from the same probe operating in fundamental collection mode, obscuring nano-scale objects of interest. In the present work, we exploit the demonstrated resolution of the second-harmonic scanning probe to identify the fundamental size limit of light localization within a disordered semiconductor film. The ~19 nm optical resolution shows the possibility of measurement of localized spots correlated to a scattering medium with an optical mean free path of ~8.8 nm. [Shafiei 2021, Lee 2018] While plasmonic light confinement and enhancement between metallic nano-junction, that comes with tunnelling breakdown threshold, has been studied via indirect measurement [Lee 2019, Savage 2012, Benz 2016], in this work we report the first direct observation of lowest spatial limit in Anderson localization of light.

**Results and Discussions:**

**1-Measuring light localization at the nanoscale**

Scattering and localization of the fundamental light (red arrows) by dislocation sites (white streaks in cross STEM) is shown in Figure 1a schematic. A 50 nm aperture fiber probe collects the SHG localized hotspots (purple color). The distance between scattering sites, the mean free path, dictates the size of the hotspots. [Shafiei 2021] Figure 1 b, c show SHG plot and cross scanning transmission electron microscope (STEM) micrograph of sample where there is no scattering sites such a dislocation inside the GaAs-GaAs III-V films, thus no light localization observed. SHG plot of GaAs-Si shows hotspots due to the presence of scattering sites that typically are larger than 200 nm in average in Figure 1d. A localized spot as small as 55 nm has been observed in this area of the sample. Analysis of this small localized light [Lee 2018, Shafiei 2021] shows that it correlates to ~14.5 nm optical MFP. This ~55 nm hotspot was the smallest SHG localized spot observed in a large statistical distribution with a variety of crystallographic defect densities. The optical resolution of the fiber probe was ~19 nm, capable of observing even smaller localized spots (SD-1).

**2-Disordered medium with ~13 nm mean free path prevents light from localizing.**

Growth parameters were modified to create III-V films filled with dislocation scattering sites with a variety of optical MFP. The sample with ~13 nm mean free path shows no sign of SHG localized spot at 390 nm (fundamental excitation at 780 nm) in Figure 1 f, g. These plots show a 5x5 μm SHG plot of $In_{30}Ga_{70}As$-GaAs with controlled mismatch and high density of dislocations and its cross STEM. No SHG hotspot is observed, signature of suppression of localized light at such an extreme confined space. The cross-section STEM images of film reveal ~13 nm average distance between dislocation all through the film. See SD-2 for more details.

**3-Access to lowest spatial limit of ~12.5 nm optical mean free path by optical breathing technique**

To confirm the lowest spatial limit with ~13 nm optical MFP, we used "breathing" of the confinement volume as incident wavelength changes and enabled a controlled tuning to the localization spatial limit. In this technique we studied the size evolution of SHG localized hotspots by very small variation of excitation wavelength to confirm the existence of the lowest spatial limit of localized light. Laser was focused on the GaAs-Si sample and scanning probe microscope collects the propagating light with filtering process that was used to discriminate the strong fundamental surface reflection and have the scattered and localized SHG light signature of scattering sites. A very small variation of excitation and collection wavelength (almost 0.01 of excited/collected λ) nudges the light from its scattering path and slightly changes constructive or destructive conditions, thus, reaching the small scattering gaps that could contribute to a new constructive or destructive condition. A large wavelength variation most likely misses many of these small scattering gaps.

Figure 2 a-g) shows multiple raster scans of the sample by varying 10 nm in excitation wavelength (5 nm in SHG collection wavelength). To study the lateral evolution of these SHG hotspots that is representing the lateral evolution of the localization of the light inside the scattering structures of threading dislocation defects, we have monitored the evolution of a specific portion of the raster scans. We plotted the 28[th] row of all the scans in Figure 2i while the SHG wavelength was changing from 420 to 390 nm. Plots in Figures-2j, k and l belong to the 100[th], 40[th] and 55[th] scanned row in Figure 2(a-g).

While Figure 2i shows very small lateral movement of localized hotspots, Figure 2j indicates that all localized hotspots exhibit some lateral movement during the wavelength alteration. This comparison helps to understand limitations of the available space between neighboring dislocation defects and leakage of light into neighboring areas under a small wavelength variation that can change the interference condition.

Disappearance of some hotspots during this $\lambda$ variation (Fig. 2j) shows that constructive conditions for interference changes and are not supported anymore by scattering sites for such a new wavelength. Figure 2l shows how reducing the wavelength causes two neighboring hotspots to merge. This is an indicator that these two scattering spaces get optically connected at this smaller wavelength.

Examining relation between changes of the localized hotspot width and wavelength, Fig.2, one can notice the hotspot size variations are distributed around a certain small number. Profile studies at these figures show that hotspots size changes are mostly in the range of ~20-60 nm by changing the collection wavelength in small steps of 5 nm. A comparison of spot 1 and 2 in Figure 2k having $\lambda$ variation between 420 nm and 415 nm, shows that hotspot width changes by ~25 nm, same as between points 2 and 3. The difference between spots 4 and 5 is ~14 nm and between spots 6 and 7 is ~21 nm. Complete comparison for all these small 5 nm SHG collected variations shows that ~ 56% of the width variations are in the range of ~20-60 nm. No width variation (~0 nm) is 17% of the total distribution, and ~130 nm variation is 18%. This observation of breathing of the hotspot mostly around ~20-60 nm width variation correlates to 12.5 nm optical mean free path. These optical breathing data show that a minimal variation in wavelength reaches the smallest gap between scattering sites. Aggregation of the localized spots around ~12.5 nm optical MFP with minimal $\lambda$ variation consistent with the existence of a lowest spatial limit for Anderson localization that discussed above.

The last image of that series, Figure 2h, demonstrates the cross STEM micrograph of the typical GaAs film that was studied in optical breathing technique. The STEM micrograph shows variety size of scattering structures.

**4-Confirming the suppression of localization of light below ~13 nm mean free path by calculation and numerical simulation**

As experimental data shows the existence of suppression of the localization of light below a specific spatial limit, we have checked the solution of Maxwell's equation for a confined area to see if the electric-field localization is suppressed below some spatial limit at specific wavelength. "V-shape" structure was chosen for such a concept to observe if electric-field localization gets suppressed and flat at some area close to the vertex where space becomes more confined. "V-shape" structure in supplementary data SD-3 is an approximation of real conditions as the wavefunction goes to zero on both structure arms as hard-walls. In real condition charges on dislocations (structure arms) repel the wavefunction but there would be leakage of electric field through them acting as soft-walls. Wavefunction and electric field in polar coordinates (Eq. 1 and 2) was derived in SD-3 and localization of light was observed in SD-4 a. Plot SD-4 b shows the profile cuts of a variety of V-shape structures with various vertex angles. In these equations, r and φ are radial distance and spherical angle in polar coordinates and $J_m(\gamma r)$ is Bessel's function in Maxwell's equation solution.

$$\psi(r, \varphi) = E_o \sin\left(\frac{n\pi}{\varphi_0}\varphi\right) J_m(\gamma r) \quad (1)$$

$$E = \pm \frac{ik}{\gamma^2}\left\{ 2\gamma \sin\left(\frac{n\pi}{\varphi_o}\varphi\right) [J_{m-1}(\gamma r) - J_{m+1}(\gamma r)] \hat{r} + \frac{1}{r}\left[\frac{n\pi}{\varphi_o} \cos\left(\frac{n\pi}{\varphi_o}\varphi\right) J_m(\gamma r)\right] \hat{\varphi} \right\} \quad (2)$$

To use more realistic conditions for experimental measurements, finite element analysis was employed to simulate the scattering and localization conditions in dislocations in disordered medium while electron charges were assigned to structural arms as defects are sink points for charges. [Shafiei 2021] We investigate the spatial limit for light localization by studying the EF localization between dislocation defect arms in V shape structures at a variety of angles. An example of confined space at a corner of structure was plotted in Figure-3 a-d for nonlinear response of the GaAs film at 390 nm. The dislocation structures were built in GaAs film while linearly polarized incoming EF have 45$^O$ incident angle at the surface of the sample (see method section for details of EF simulation). Plots in 3(a-d) show dislocation structure with 74, 37, 24 and 15 $^O$ angle. Localized EF gets pushed away from the corner of the structure when the angle becomes smaller and EF becomes more flat in the space at the corner of the structure close to the vertex. We have studied the EF-profile cuts of many V-shape scattering structures, such shown in Figure 3e. Figure 3b shows one of these profile cuts starting at the corner of V-shape structure and through the localized EF between dislocation arms of the V structure. Profile cuts in Figure 3e show that in larger corner angle structures (74$^O$), localized EF can reach the V structure vertex. When V-shape structures become tighter, localized EF confined at corner, get suppressed and pushed away. By decreasing the corner angle of structure to 15$^O$, EF profile does not show localization at the corner of structure. Here the flat EF is an order of magnitude smaller than the localized EF for this angle and extended up to ~50 nm away from the corner of the structure. The simulation shows that suppressed light is very weak but not zero. Confined space is shielded from most of the scattered light reaching the corner of the V-shape structure. Enough scattering light that is required for interference pattern in Anderson localization of light [John 1991], does not reach to that confined space and electric-fields localization get suppressed in that vicinity.

This is a signature of the existence of a fundamental size limit for localized light. The EF gets suppressed in areas with ~12.6 nm opening width (marked by an arrow on SHG plot Figure 3 (d) and profile cut (e)). This supports our observation of suppression of scattering and localization of light in disordered mediums with ~13 nm mean free path (Figure 1 f and g). Further decrease of the corner angle of scattering V-shape structure ($10^O$) leads to additional space confinement causing the extent of suppression of light localization. Detailed data of suppression of light localization with flat and weak EF at confined space are shown in Supplementary Data SD-5.

Supplementary Data SD-6, shows Simulation of SHG reflection and scattering of light in the Air-GaAs-Si medium where the thin film is filled with low random scattering dislocation. Disordered medium with ~20 nm dislocation mean free path (average gap between dislocations) in SD-6 (a) shows that light still penetrates inside the film and gets localized between the dislocations. When the average gap distance between dislocation becomes ~12 nm (SD-6 (b)), SHG hotspot disappears as that light could not localize inside the film and instead get scattered off of the scattering medium. Like previous data, incoming light is at 780 nm and SHG collection (plotted data) is at 390 nm.

We have plotted the plasmonic enhancement field at a similar V-shape structure made of silver medium and air V-shape cone in comparison to some of the localized EF from dislocation V-shape structure (SD-7). The plot shows how the EF plasmonic enhancement is different from light Anderson localization at the corner of V-shape structure and no localization exists at the vertex area. Plasmonic enhancement plot is plotted at fundamental light (780 nm) with intensity around $10^{34}$ V/m while SHG (390 nm) localized spots are in order of $10^{-12}$ V/m.

## 5-Lowest spatial limit for light scattering and Anderson localization

Figure 4 shows the localization lengths (hotspots sizes) measured in variety of III-V films with different dislocation defect density (variety of optical mean free path) that are linked to the spacing limit where Anderson localization get suppressed. Selected experimental localization length of SHG hotspots (in trigonometric symbols) are plotted as a function of optical mean free path (gap between scattering dislocations). The mean free paths for these individual data are calculated from the hotspot sizes (from Anderson localization theory). [Lee 2018, Shafiei 2021] The averages of these localization lengths are plotted for different films (in black triangles with error bars). This includes the smallest localization length of ~ 55 nm observed in Figure- 1d (correlated to ~14.5 nm mean free path). The very small localized spots become less frequent. Probability distribution of measured localized spot as function of optical MFP (not shown here) shows that as we get close to the fundamental spatial limit of localized spot, there is a very low number of observed spots. The perpendicular red line identifies the suppression of the Anderson localizations observed experimentally in the film with ~13 nm optical mean free path

(Figure-1 f,g). The gray line shows the lowest limit of Anderson localization length that was observed in simulations of the V-shape scattering structure of ~12 nm width (Figure 3). The ~ 19 nm resolution of a fiber based scanning probe microscope is marked by a blue line, corresponding to about 8.8 optical mean free path (Figure SD-1). The experimental and simulation data shows that with the sufficient optical resolution of the probe, we do not observe light Anderson localization below corresponding ~13 nm mean free path at the studied wavelength.

**Conclusion:**

Fundamental lowest spatial limit of localization of light has been directly observed through the study of light scattering in disordered media using a high resolution fiber-based scanning probe microscope. Threading dislocations offering a wide range of continuous distribution of intra-defects gap sizes have been used as the scattering sites. The study shows that a disordered medium with ~13 nm mean free path prevents localization of light (SHG observation λ ~ 390 nm), pointing to the existence of such a critical gap size of suppressed localization. Experimental and calculation observation of suppression of localization of light at such a scale, clarifies the details of transition of electromagnetism fields from diffusion to localization and suppression at nano-scale.

**Methods:**
**Experimental Setup and Measurement Technique:**

Second-harmonic probe microscope (SHPM) was used to collect the SHG signal through a fiber based nearfield scanning optical microscope (NSOM) with extreme sub-wavelength resolution. [ Shafiei 2022, Shafiei 2021 ] A 76 MHz laser with ~150 fs pulse width at ~780 nm was focused on an area about ~10 μm on the sample. Collection of SHG was done at 390 nm through an uncoated fiber probe with 50 nm aperture. The uncoated probe was preferred over coated probe to avoid and minimize any light and electromagnetic field enhancement and perturbation at the probe area. The uncoated probe was kept at ~ 20-30 nm above the sample by using a feedback loop system monitoring the amplitude of the vibrating probe during the scanning process.
The incident laser excitation angle was ~45° with linear polarization. The Sample was scanned by a piezoelectric stage under the stationary probe. A combination of fiber, filters and Photomultiplier tube (PMT) was used to reject dominating fundamental light and has a noise free SHG result. This was necessary as nonlinear response of the GaAs-Si film is typically ~$10^{15}$ times weaker than the fundamental light. Fundamental light was scattered by subsurface crystallographic defects and get localized and crated hotspots. These weak hotspots are not detectable at dominating surface fundamental reflection but can be distinguished by their SHG signature from the film typical SHG response. The resolution power of the scanning probe microscope was explored and confirmed (SD-1). Details of setup, collection of the signal and procedure of pulling and preparation of 50 nm aperture probe are explained in our previous publications. [ Shafiei 2022, Shafiei2021 ]

**Numerical Calculation :**

Finite element analysis was used by using COMSOL software to complete the calculation of electric field (optical intensity study) inside and outside the III-V materials. Electromagnetic Wave Frequency Domain package was used to calculate near-field details of electromagnetic fields in addition to far-field features. In Fig.3 and SD-6 and 7 linear polarized 780 nm excitation field propagating from left side and make incident on GaAs sample that has been arranged at 45⁰ respect to horizontal axes. This replicates the configuration of excitation of sample and light collection. SHG 390 nm response of the film (along with fundamental response) were calculated and plotted. The fundamental and second harmonic frequency domains are coupled through polarization definitions using electric field components. $\sim 1\times 10^{-22}$ F/V was used as a nonlinear matrix element for GaAs. The left boundary included the incoming linear electric field and the other boundaries are free of any field and reflection. Scattering properties of dislocation defects were replicated by assigning electron charges to these defects based on previous experimental and theoretical publications showing that dislocations act as sink point for electron. All the plots display averaged electric fields (average over all directions) as scalar elements. In Fig. 3 V-shape structure as scattering structure was prepared as explained in text as threading dislocation. The scattering V-structure is within the 500 nm thick GaAs film and has been studied with variety of orientations and configuration to measure the suppression of Anderson localization of light.

**Sample Growth:**

III-V sample films were grown over Si and Ge substrate under variety of condition to control the growth of threading dislocations and distances between these scattering sites down to few nm range. The III-V film such as GaAs was grown on on-axis Si (001) by metal-organic chemical vapor deposition (MOCVD) technique using two-step growth approach. AIXTRON CRIUS-R MOCVD system was used for that purpose. Essential silicon wafer cleaning and hydrogen passivation was done by vapor HF and wet HF processes. To promote the formation of double steps on Si along <110> direction for prevention of antiphase domains, baking at high temperature (>800 C) was performed. III–V films were grown by using trimethylindium (TMIn) and trimethylgallium (TMGa) as the group-III precursors, tertiarybutylarsine (TBAs), tertiarybutylphosphine (TBP) as well as arsine ($AsH_3$) and phosphine ($PH_3$) as the group V precursors. To have charge neutrality along the interface and promote the growth of single domain GaAs, wafer surface was saturated with an arsenic monolayer by introducing TBAs in the reactor at low temperature (<500 C). Two step growth was introduced by <20 nm GaAs LT nucleation layer at (<450 C) by low V/III ration with roughness ~1nm measured by AFM and SEM. A 500nm thick GaAs was grown at ~600 C by using $AsH_3$ with high V/III ratio and growth rate of ~1.3 micrometer/h with ~0.6nm roughness. Quality and defect density of the crystal was checked by a high-resolution X-ray diffraction (HRXRD) and cross and planar STEM later. Annealing had performed at the end to improve the quality of the crystal ay 750C. Details are discussed in previous publication .[Shafiei 2021]

Alternative III-V and SiO2 stripe pattern ART sample with flat surface was prepared to create sharp optical edges. These optically sharp edges were used characterized the uncoated fiber probe performance and resolution (SD-1). An AIXTRON CRIUS-R MOCVD system was used to grow InP films on patterned 300mm wafers. On-axis Si

(001) wafers were used to fabricate experimental test structures. These structures were created by forming a 180 nm thick thermal $SiO_2$ layer followed by photo lithography and dry etching of trenches in the oxide in the [110] direction. 65 nm wide trenches were opened in the oxide and spaced at 130 nm pitch. The dimension of the trenches along the [110] direction was 25.4 mm. Prior to the epitaxial growth of InP, the residual and native oxide at the bottom of the trenches was pre-cleaned and removed by a vapor $HF/NH_3$ process. For the growth of InP films, trimethylindium (TMIn) was used as the group-III precursor, and tertiarybutylphosphine (TBP) and phosphine ($PH_3$) were used as the group V precursors. The growth was carried out at low pressure. A two-step growth approach was implemented, in which the first step aimed at depositing an InP seed layer at low temperature (below 425 °C) using TBP with a V/III ratio of 25, and the second step is needed to bulk fill the oxide trenches with InP at 600 °C and with a V/III ratio of 100 by utilizing $PH_3$ precursor.

**Acknowledgments:** The work was supported in part by the Welch Foundation grant.

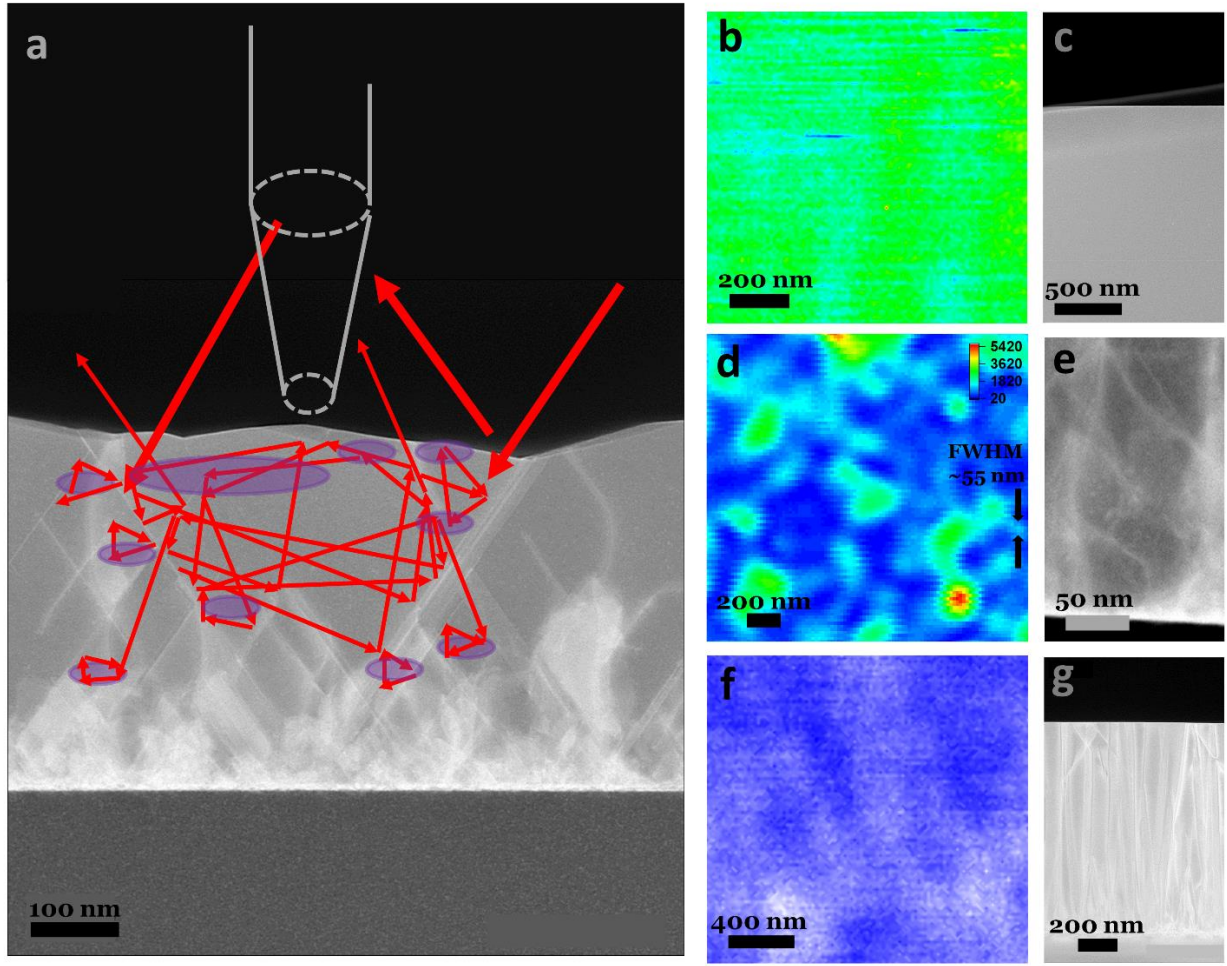

**Figure 1 | Optical Anderson localization in disordered medium correlated to distance between scattering sites (mean free path). (a)** Schematic of light scattering and localization of fundamental light (red arrows) by scattering dislocation sites (white streaks) and collection of overall SHG localization signature (purple color) by 50 nm aperture fiber probe. **(b, c)** SHG plot with no localization and its corresponding Cross-x STEM of GaAs-GaAs film without any dislocations. **(d, e)** SHG plot with average +200 nm hotspots and smallest localization length hotspots of ~55 nm and its corresponding Cross-x STEM of GaAs-Si with moderate density of dislocations. **(f, g)** SHG plot with no light localization and its corresponding Cross-x STEM of $In_{30}Ga_{70}As$-GaAs full of dislocations with ~13 nm average distance between scattering dislocations (mean free path).

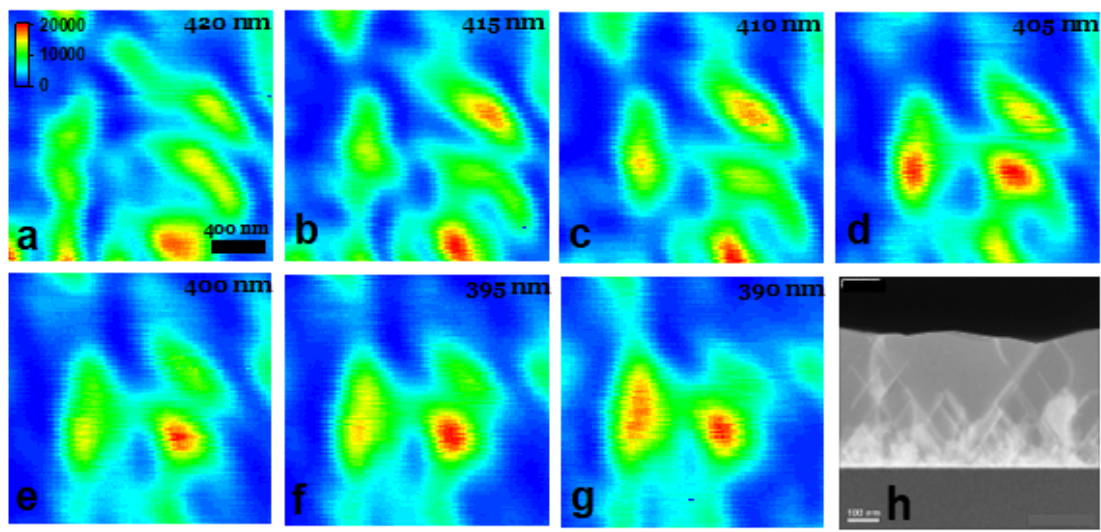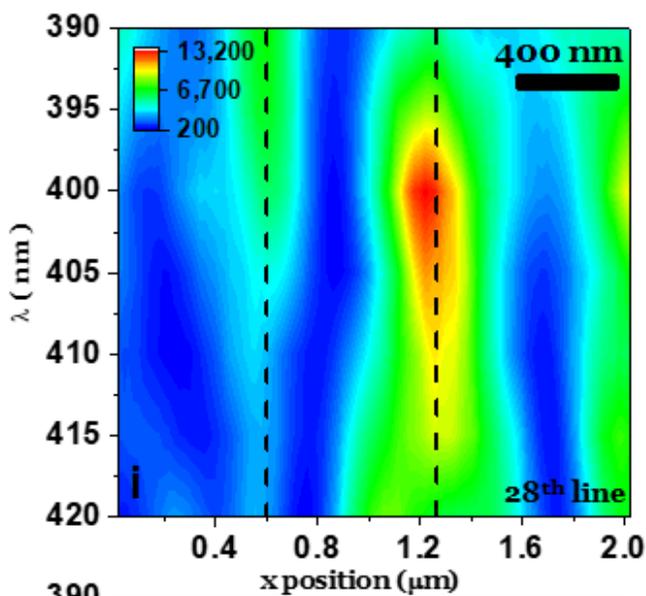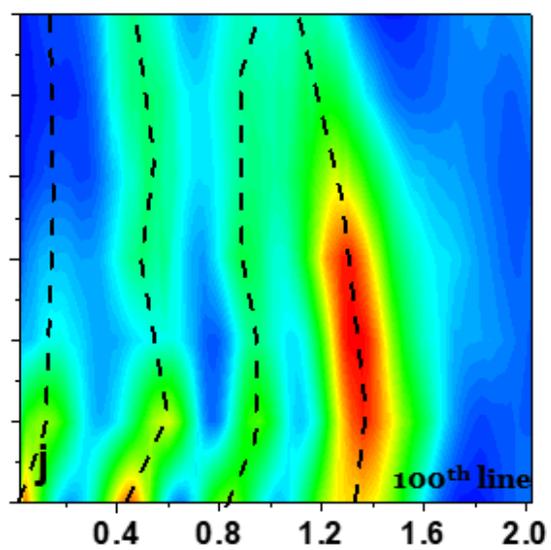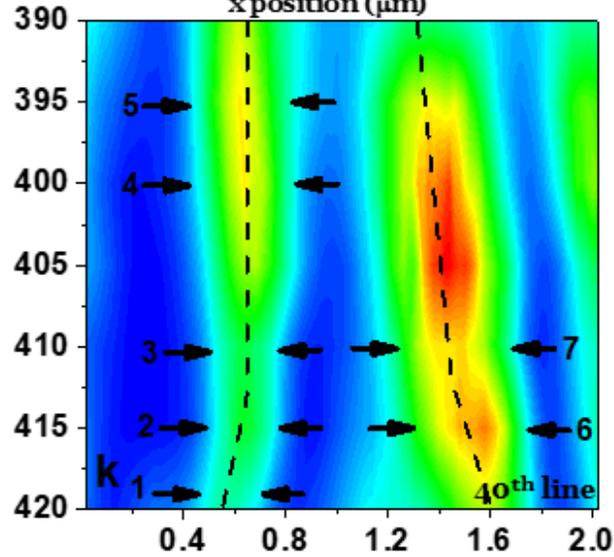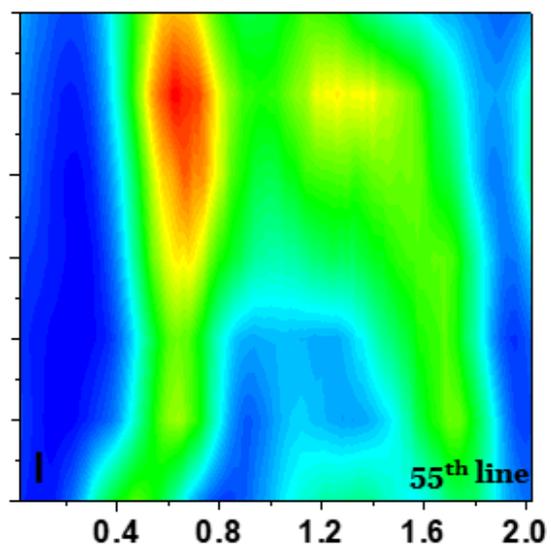

**Figure 2 | "Optical breathing": evolution of light localization by small wavelength variations, reaching small gaps between dislocation scattering structures. (a-g)** SHG raster scan plots of GaAs-Si film by a fiber probe when the collection wavelength λ changes in the range of 420 nm to 390 nm in 5 nm step (excitation pulsed laser is changing from 840 nm to 780 nm in 10 nm step). **(h)** Cross STEM micrograph of GaAs-Si film shows a variety of size of scattering structures. **(i-l)** Profile-cut along specific row of plots (a-g), demonstrating the variation (optical breathing) of localized spot widths with the varying λ. Plot **(i)** shows the 28$^{th}$ row of all variation throughout the entire λ range. Plot **(j)**, **(k)** and **(l)** belong to the 100$^{th}$, 40$^{th}$ and 55$^{th}$ scanned row of plots (a-g). Hotspot widths were monitored by comparing their width at neighboring wavelengths. The arrows in **(k)** display some of these comparison points. The comparison shows that small variations in wavelength cause the majority of width variations to be in the range of ~20-60 nm with average optical mean free path of ~12.5 nm. Such a small variation in wavelength (5 nm in SHG measurements) accesses the smallest gaps between scattering sites defining the lowest spatial limit. Disappearance of some of the hotspots are clear in some of the last plots (i-l) by changing the wavelength and SHG plot in (l) shows the merging of two neighboring localized hotspots, caused by decreasing wavelength from 420 nm to 390 nm.

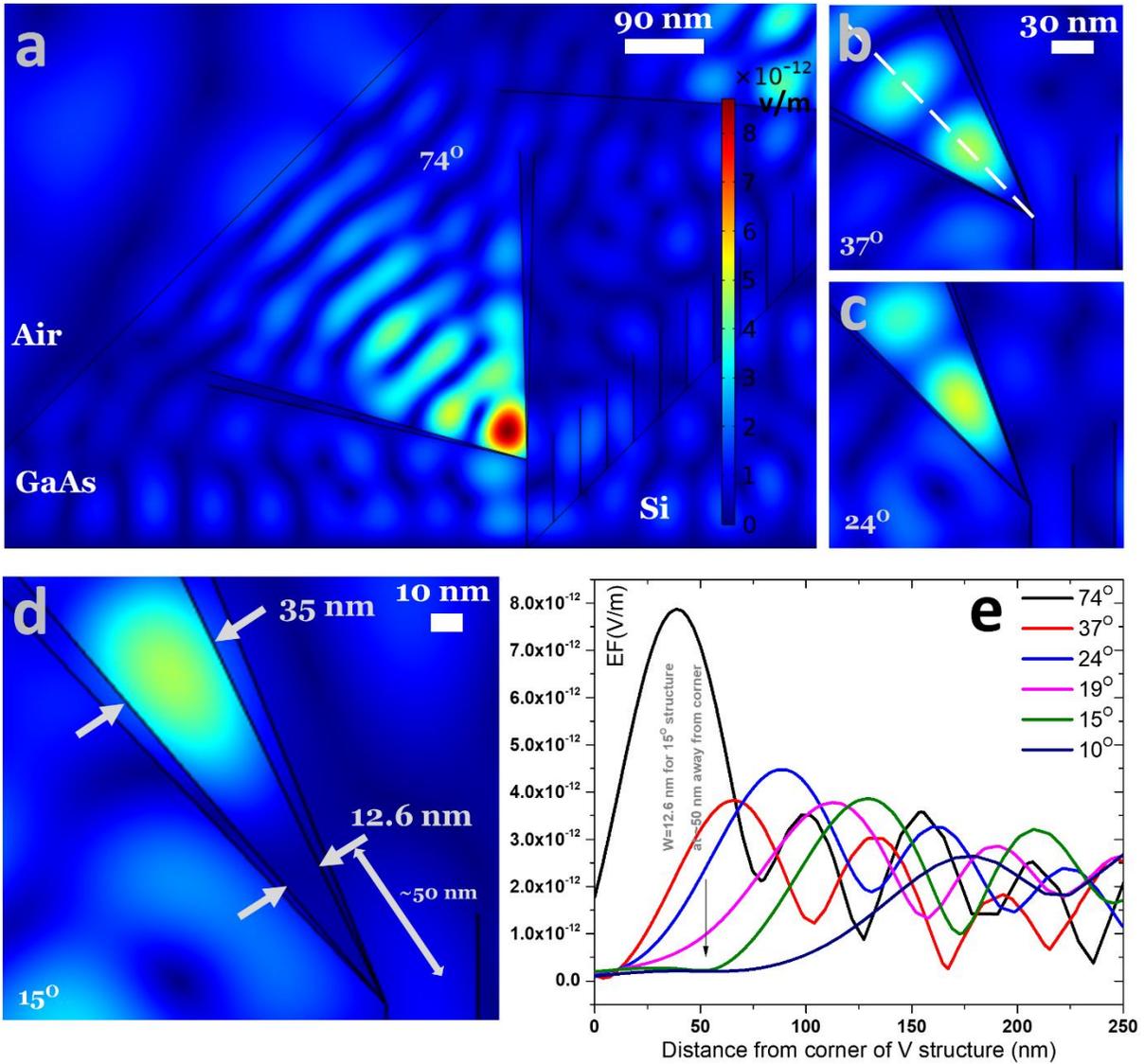

**Figure 3 | Simulation of light (electric field) shows the suppression of Anderson localization in a confined space in the V-shape dislocation structure. (a-d)** Localized light in SHG regime (390 nm) gets gradually suppressed in V-shape dislocation scattering structure as vertex angle decreases in GaAs film. **(e)** Profile cuts of EF intensity in confined space inside the V-shape scattering structures of different vertex angles show such a suppression at vertex area. EF localization is observed at 74° angle (black line) and it gets fully suppressed down at 15° structure (green line) at corner space of vertex. This suppressed EF extends from vertex up to ~50 nm away from the corner of the structure where the structure has ~12.6 nm opening (marked in d and e). Suppression of localized EF at an area with ~12.6 nm opening width defines the lowest spatial limit for light Anderson localization of this width (d).

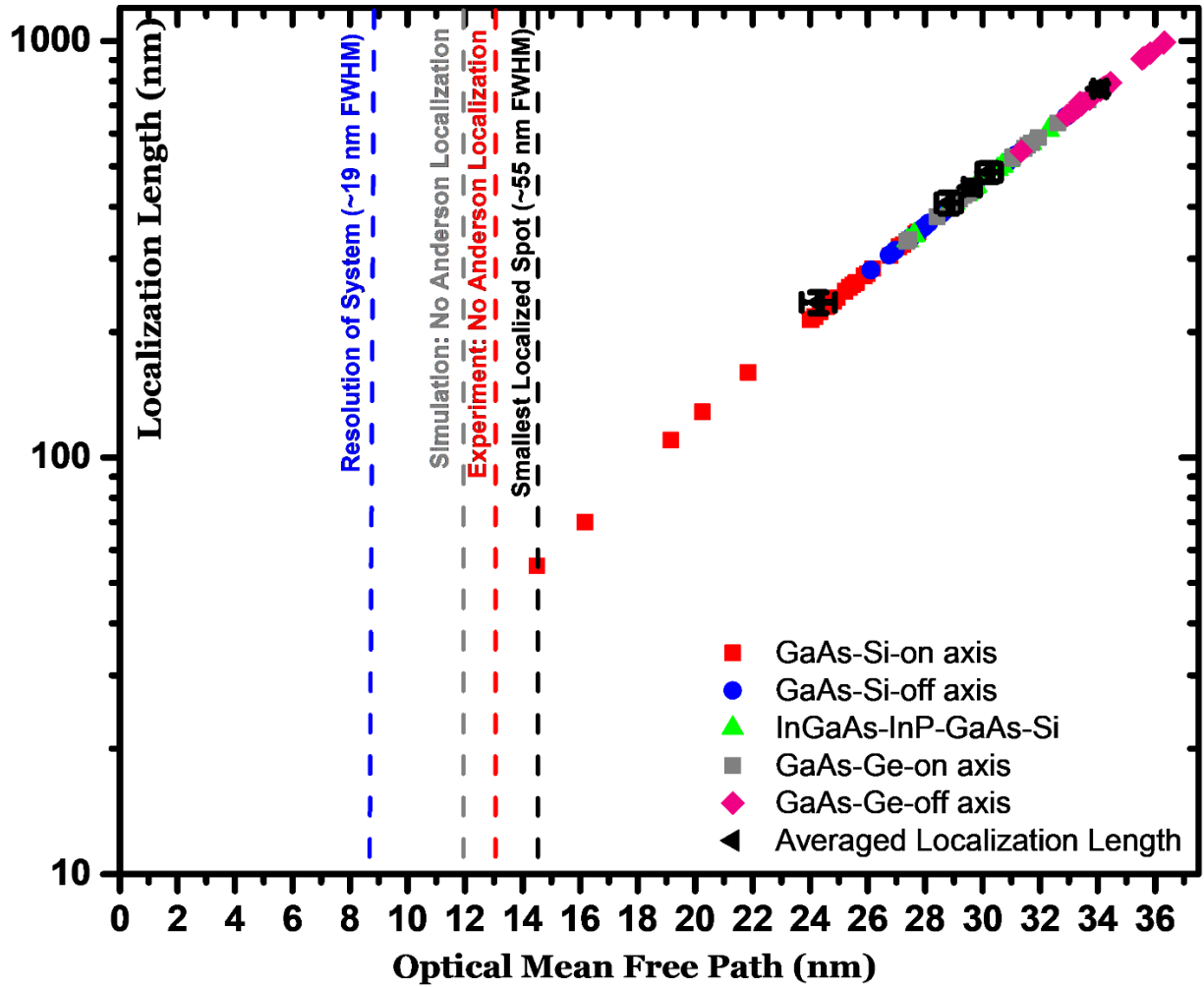

**Figure 4 | Observation of lowest spatial limit of light Anderson localization by measurements and simulations.** Light localization length (size of hotspots) measurements in films with different dislocation defect density (different optical mean free path) are plotted in different trigonometric symbols for a set of listed material stacks. Optical MFPs are calculated based on hotspot sizes. Averages of these localization lengths are plotted for each investigated III-V film in black traingles with error bars. The very small localized spots become less frequent as we get close to the observed limitation. The smallest localized spot in a large statistical distribution is marked by black line (with ~14.5 nm MFP)(from Fig. 1d). The experimental suppression of Anderson localization in the sample with ~13 nm optical MFP is shown in the red vertical line (from Fig. 1 f and g). The simulated lowest limit of Anderson localization in gray line with ~12 nm gap is in agreement with experimental observations (from Fig. 3, SD-5 and SD-6). ~19 nm resolution of the probe microscope in blue line (correlated to 8.8 optical MFP/ from Fig. SD-1) shows that the system is capable of measuring smaller light localized length.

# References

[John 1991] John, Sajeev. "Localization of light." *Physics Today* 44, 32 (1991

[Wiersma 1997] Wiersma, Diederik S., Paolo Bartolini, Ad Lagendijk, and Roberto Righini. "Localization of light in a disordered medium." *Nature* 390, 671 (1997)

[Shafiei 2021] Shafiei, F., Orzali, T., Vert, A., Miri, M.A., Hung, P.Y., Wong, M.H., Alù, A., Bersuker, G. and Downer, M.C., 2020. Detection of Subsurface, Nanometer-Scale Crystallographic Defects by Nonlinear Light Scattering and Localization. *Advanced Optical Materials* 9, 2002252 (2021).

[Wiersma 2013] Wiersma, D. S. Disordered photonics. *Nat. Photonics* 7, 188-196 (2013).

[Sapienza 2010] Sapienza, L., Thyrrestrup, H., Stobbe, S., Garcia, P.D., Smolka, S. and Lodahl, P., 2010. Cavity quantum electrodynamics with Anderson-localized modes. *Science*, *327*(5971), pp.1352-1355.

[Segev 2013] Segev, M., Silberberg, Y., Christodoulides, D. N. Anderson localization of light. *Nat. Photonics* 7, 197-204 (2013).

[Wang 2020] Wang, P., Zheng, Y., Chen, X., Huang, C., Kartashov, Y.V., Torner, L., Konotop, V.V. and Ye, F., 2020. Localization and delocalization of light in photonic moiré lattices. *Nature*, *577*, 46 (2020)

[Cao 1999] Cao, H., Zhao, Y.G., Ho, S.T., Seelig, E.W., Wang, Q.H. and Chang, R.P., Random laser action in semiconductor powder. *Physical Review Letters*, *82*, 2278 (1999)

[Sperling 2013] Sperling, T., Buehrer, W., Aegerter, C. M., Maret, G. Direct determination of the transition to localization of light in three dimensions. *Nat. Photonics* 7, 48-52 (2013).

[Schuurmans 1999] Schuurmans Frank, J.P., Vanmaekelbergh, D., van de Lagemaat, J. and Lagendijk, A., Strongly photonic macroporous gallium phosphide networks. *Science*, *284*, 141 (1999)

[Mascheck 2012] Mascheck, M., Schmidt, S., Silies, M., Yatsui, T., Kitamura, K., Ohtsu, M., Leipold, D., Runge, E. and Lienau, C., Observing the localization of light in space and time by ultrafast second-harmonic microscopy. *Nature Photonics* **6**, 293 (2012).

[Smolyaninov 1997] Smolyaninov, I.I., Zayats, A. V., Davis, C. C. Near-field second harmonic generation from a rough metal surface. *Phys. Rev. B* **56**, 9290-9293 (1997).

[Hofling 2021] Höfling, M., Zhou, X., Riemer, L.M., Bruder, E., Liu, B., Zhou, L., Groszewicz, P.B., Zhuo, F., Xu, B.X., Durst, K. and Tan, X. Control of polarization in bulk ferroelectrics by mechanical dislocation imprint. *Science*, *372*, 961 (2021)

[Kim 2015] Kim, S.I., Lee, K.H., Mun, H.A., Kim, H.S., Hwang, S.W., Roh, J.W., Yang, D.J., Shin, W.H., Li, X.S., Lee, Y.H. and Snyder, G.J. Dense dislocation arrays embedded in grain boundaries for high-performance bulk thermoelectrics. *Science*, *348*, 109 (2015)

[Dam 1999] Dam, B., Huijbregtse, J.M., Klaassen, F.C., Van der Geest, R.C.F., Doornbos, G., Rector, J.H., Testa, A.M., Freisem, S., Martinez, J.C., Stäuble-Pümpin, B. and Griessen, R. Origin of high critical currents in YBa$_2$Cu$_3$O$_{7-\delta}$ superconducting thin films. Nature, 399, 439 (1999)

[Yang 2019] Yang, Y., Zhu, D., Yan, W., Agarwal, A., Zheng, M., Joannopoulos, J.D., Lalanne, P., Christensen, T., Berggren, K.K. and Soljačić, M. A general theoretical and experimental framework for nanoscale electromagnetism. *Nature*, *576, 248* (2019)

[Rotenberg 2014] Rotenberg, N. and Kuipers, L. Mapping nanoscale light fields. *Nature Photonics 8,* 919 (2014)

[Savage 2012] Savage, Kevin J., Matthew M. Hawkeye, Rubén Esteban, Andrei G. Borisov, Javier Aizpurua, and Jeremy J. Baumberg. "Revealing the quantum regime in tunnelling plasmonics." *Nature* 491, 574 (2012)

[Lee 2018] Lee, M., Lee, J., Kim, S., Callard, S., Seassal, C. and Jeon, H., 2018. Anderson localizations and photonic band-tail states observed in compositionally disordered platform. *Science advances*, *4*(1), p.e1602796.

[Anderson 1958] Anderson, P.W., Absence of diffusion in certain random lattices. *Physical Review*, *109,* 1492 (1958)

[Lagendijk 2009] Lagendijk, A., van Tiggelen, B. A., Wiersma, D.S. Fifty years of Anderson localization. *Phys. Today* 62, 24-29 (2009).

[Mosk 2012] Mosk, A.P., Lagendijk, A., Lerosey, G. and Fink, M. Controlling waves in space and time for imaging and focusing in complex media. *Nature photonics*, *6*, 283 (2012)

[Willig 2006] Willig, Katrin I., Robert R. Kellner, Rebecca Medda, Birka Hein, Stefan Jakobs, and Stefan W. Hell. "Nanoscale resolution in GFP-based microscopy." *Nature Methods* 3, 721 (2006)

[Betzig 2006] Betzig, Eric, George H. Patterson, Rachid Sougrat, O. Wolf Lindwasser, Scott Olenych, Juan S. Bonifacino, Michael W. Davidson, Jennifer Lippincott-Schwartz, and Harald F. Hess. "Imaging intracellular fluorescent proteins at nanometer resolution." *Science* 313, 1642 (2006)

[de Aguiar 2017] de Aguiar, H.B., Gigan, S. and Brasselet, S., Polarization recovery through scattering media. *Science Advances*, *3*(9), p.e1600743 (2017)

[Ritcher 2018] Richter, Marten, Rohan Singh, Mark Siemens, and Steven T. Cundiff. "Deconvolution of optical multidimensional coherent spectra." *Science advances* 4 (6) eaar7697 (2018)

[Stephens 2003] Stephens, David J., and Victoria J. Allan. "Light microscopy techniques for live cell imaging." *Science* 300, 82 (2003)

[Feng 2019] Feng, Lei, Jiazhong Hu, Logan W. Clark, and Cheng Chin. "Correlations in high-harmonic generation of matter-wave jets revealed by pattern recognition." *Science* 363, 521 (2019)

[DeMarco 2021] DeMarco, B. and Thywissen, J.H., No vacancy in the Fermi sea. *Science 374*, 936 (2021)

[Shafiei 2022] Shafiei, Farbod, and Michael C. Downer. "Collection of Propagating Electromagnetic Fields by Uncoated Probe." *Ultramicroscopy* 240, 113597 (2022).

[Lee 2019] Lee, Joonhee, Kevin T. Crampton, Nicholas Tallarida, and V. Ara Apkarian. "Visualizing vibrational normal modes of a single molecule with atomically confined light." *Nature* 568, 78 (2019)

[Benz 2016] Benz, Felix, Mikolaj K. Schmidt, Alexander Dreismann, Rohit Chikkaraddy, Yao Zhang, Angela Demetriadou, Cloudy Carnegie et al. "Single-molecule optomechanics in "picocavities"." *Science* 354, 726 (2016)

**Supplementary Data**

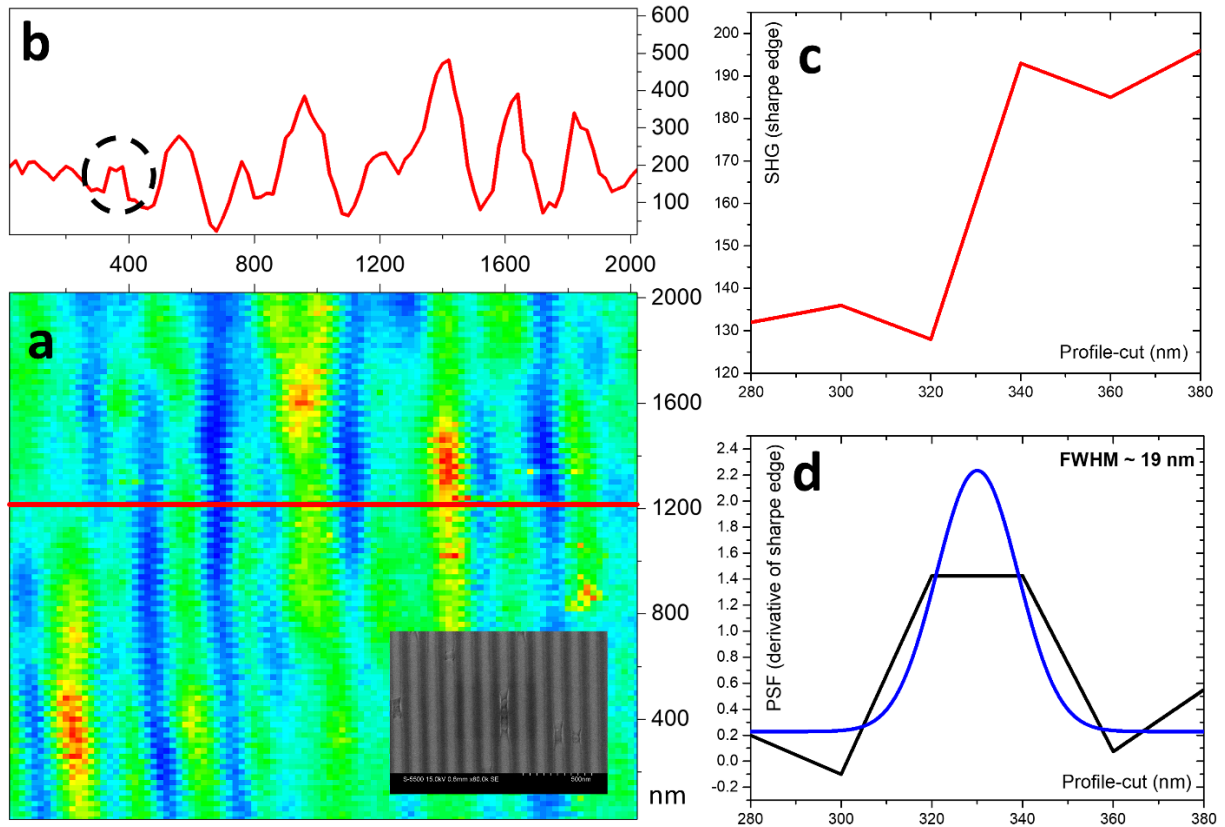

**Supplementary Data Fig. SD1 ) Optical resolution of fiber based scanning probe system with 50 nm aperture:**
**(a)** A stripe pattern of alternative InP and SiO2 (each stripe 70 nm wide) was used to have a very sharp optical edge. The 2x2 μm sample was scanned by a 50 nm uncoated probe and SHG signal was collected. Inset shows the electron micrograph of the polished InP-SiO2 stripe sample. InP with a strong source of SHG signal and SiO2 with an almost dim source of such a signal, creates a sharp optical edge. A profile-cut at y~1200 nm was sown in **(b)**. A marked sharp edge in (b) was picked and plotted individually (the left side of the step pattern **(c)**. A derivative of this sharp edge was plotted **(d)** representing the point spread function of the smallest optical feature with roughly 19 nm FWHM. This represents the optical resolution of the 50 nm aperture uncoated fiber probe.

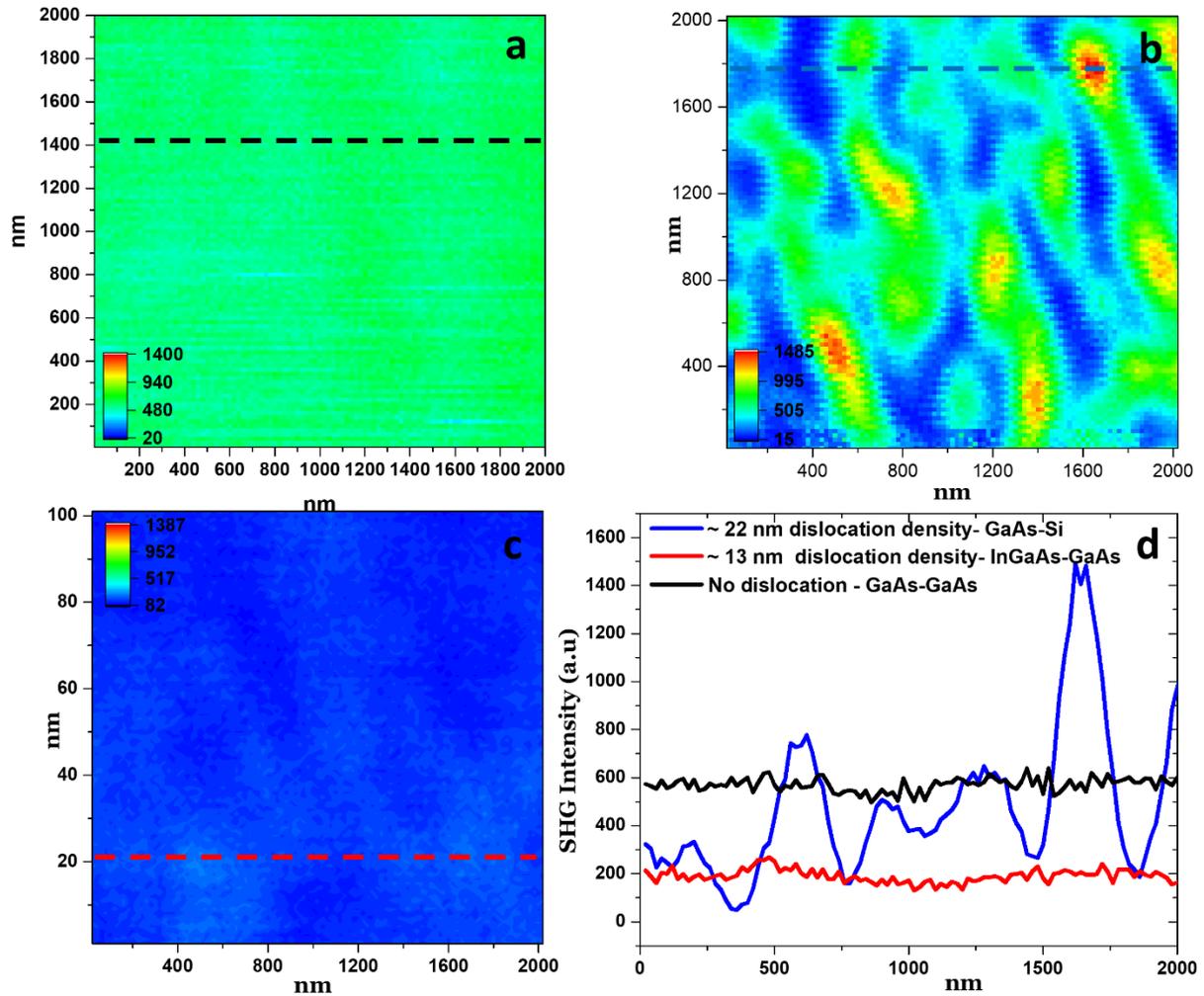

**Supplementary Data Fig. SD2 ) Comparison of SHG profile-cut for films with different dislocation density:**
**(a)** GaAs-GaAs with no defect shows no localization spots **(b)** GaAs-Si with ~24 nm dislocation gap (mean free path) shows localization spots with + 200 nm FWHM. **(c)** InGaAs-GaAs sample with ~13 nm dislocation gap (mean free path) shows no localization and suppressed intensity **(d)** Profile-cuts for these 3 different dislocation gaps samples are shown here. Presence of dislocations in the film (comparing a and b) causes scattering and creation of localized spots. Later if the density of dislocation increase to the level that the dislocation gap (optical mean free path) is smaller than ~13 nm, localization of light gets suppressed (comparing b and c). Detailed comparison and signature of light localization has been described in previous publication. [Shafiei2021]

## Supplementary Data SD3) Calculation of localized EF and its fundamental size limit by solving Maxwell's equation in V-shape confined space:

We have solved Maxwell's equation for a model consisting of "V-shape" structure shown in Fig. 1. with initial angle φ₀. Maxwell's equations in a region with zero charge and current have been defined.

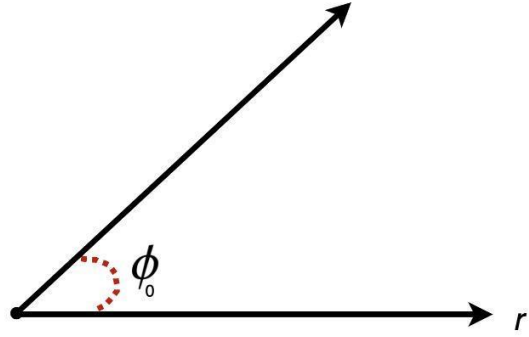

Figure 1: The two dimensional cut of the region.

In this case the Maxwell's equations reduce to:

$$\nabla \cdot \mathbf{E} = 0$$
$$\nabla \cdot \mathbf{B} = 0$$
$$\nabla \times \mathbf{E} = \frac{\partial B}{\partial t}$$
$$\nabla \cdot \mathbf{B} = -\mu \varepsilon \frac{\partial E}{\partial t} \quad (1)$$

Using the ∇× we can obtain the following equations:

$$(\nabla^2 - \mu \varepsilon \frac{\partial}{\partial t}) \ \{E \ B\} = 0 \quad (2)$$

Assuming wave can travel freely in z–direction as $e^{\pm ikz - i\omega t}$ we find the transverse part as following:

$$\nabla^2 + (\mu\varepsilon\omega^2 - k^2) \{E(x,y) \ B(x,y)\} = 0 \quad (3)$$

The transverse fields for TM waves and TE waves can be defined as:

$$\{E(x,y) \ B(x,y)\} = \pm \frac{ik}{\gamma^2} \nabla_t \psi \quad (4)$$

Here $\gamma^2 = \mu\varepsilon\omega^2 - k^2$ which result in:

$$(\nabla^2_t + \gamma^2) \psi = 0 \quad (5)$$

with the following boundary conditions as we consider simple case of hard walls and no charges on the V-shape structure and other places:

$$\{\psi|_s = 0 \text{ or } \partial\psi \, \partial n|_s = 0 \tag{6}$$

For the problem with such symmetry its useful to use cylinder coordinate as follows:

$$1/r \frac{\partial}{\partial r}(r \frac{\partial \psi}{\partial r}) + 1/r^2 \frac{\partial^2 \psi}{\partial \varphi^2} + (\mu\varepsilon\omega^2 - k^2)\psi = 0 \tag{7}$$

with solution of $\varphi$ part as $(A \sin(m\varphi) + B \cos(m\varphi))\rho(r)$. Since we are expecting $\psi = 0$ at $\varphi = 0$ only $\sin(m\varphi)$ will survive. Moreover $\psi = 0$ at $\varphi = \varphi_0$ as well which will result in $m\varphi_0 = n\pi$ or $m = n\pi/\varphi_0$. Next we are going to solve the r-dependent part. We can re-write equation (3) as:

$$1/r \frac{\partial}{\partial r}(r \frac{\partial \rho(r)}{\partial r}) - m^2/r^2 \frac{\partial^2 \psi}{\partial \varphi^2}\rho(r) + (\mu\omega^2 - k^2)\rho(r) = 0 \tag{8}$$

or

$$r^2 \partial^2\rho(r)/\partial r^2 + r\, \partial\rho(r)/\partial r + [(\mu\varepsilon\omega^2 - k^2)r^2 - m^2]\rho(r) = 0 \tag{9}$$

defining $x = \sqrt{(\mu\varepsilon\omega^2 - k^2)}\, r = \gamma r$ we can simplify this equation as:

$$x^2 \partial^2\rho(r)/\partial x^2 + x\, \partial\rho(r)/\partial x + [x^2 - m^2]\rho(r) = 0 \tag{10}$$

Because we expect the $\psi$ to be finite at $r = 0$ the only accepted solution is the $J_m(x)$. So the wavefunction solution is:

$$\psi(r, \varphi) = A \sin\left(\frac{n\pi}{\varphi 0}\varphi\right) J_m(\gamma r) \tag{11}$$

Now we find the maximums of the $\psi(r, \varphi)$ in r direction

$$\frac{d\psi(r,\varphi)}{dr} = A \sin\left(\frac{n\pi}{\varphi 0}\varphi\right)\gamma \frac{d\, Jm(\gamma r)}{dr} = A \sin\left(\frac{n\pi}{\varphi 0}\varphi\right) 2\gamma\, (J_{m-1}(\gamma r) - J_{m+1}(\gamma r)) = 0 \tag{12}$$

$$J_{m-1}(\gamma r) = J_{m+1}(\gamma r) \tag{13}$$

For m = 0 we find that

$$J_{-1}(\gamma r) = J_1(\gamma r) \rightarrow J_1(\gamma r) = 0 \tag{14}$$

which is basically all the roots of Bessel function $J_1(\gamma r)$. For n = 1 we plotted a profile-cut of a variety of angles $\varphi_0 = 5^O, 10^O, 13^O, 15^O, 19^O, 24^O$. Here assuming k = 0 we find $\gamma = \sqrt{\varepsilon\mu}\omega$. Using $c = n/\sqrt{\varepsilon\mu}$ where c is the speed of light and n is the refraction index. we find that: $\gamma = n\omega/c = 2\pi nc/\lambda c = 2\pi/\lambda$ in this experiment. The position of the first peak/Max of plot $\varphi_0 = \sim 15^O$ which was the best fit for experimental results was checked as well.

Electric-field then is calculated from wavefunction:

$$E = \pm \frac{ik}{\gamma^2} \nabla \psi (r, \varphi) \tag{15}$$

$$E = \pm \frac{ik}{\gamma^2} \left\{ \frac{\partial}{\partial r} \hat{r} + \frac{1}{r} \frac{\partial}{\partial \varphi} \hat{\varphi} \right\} \psi (r, \varphi) \tag{16}$$

$$\tag{17}$$

$$E = \pm \frac{ik}{\gamma^2} \left\{ 2\gamma \sin\left(\frac{n\pi}{\varphi_o} \varphi\right) [J_{m-1}(\gamma r) - J_{m+1}(\gamma r)] \hat{r} + \frac{1}{r} \left[\frac{n\pi}{\varphi_o} \cos\left(\frac{n\pi}{\varphi_o} \varphi\right) J_m(\gamma r)\right] \hat{\varphi} \right\}$$

While the intensity of E is given by $E = \sqrt{|E_r|^2 + |E_r|^2}$ \qquad (18)

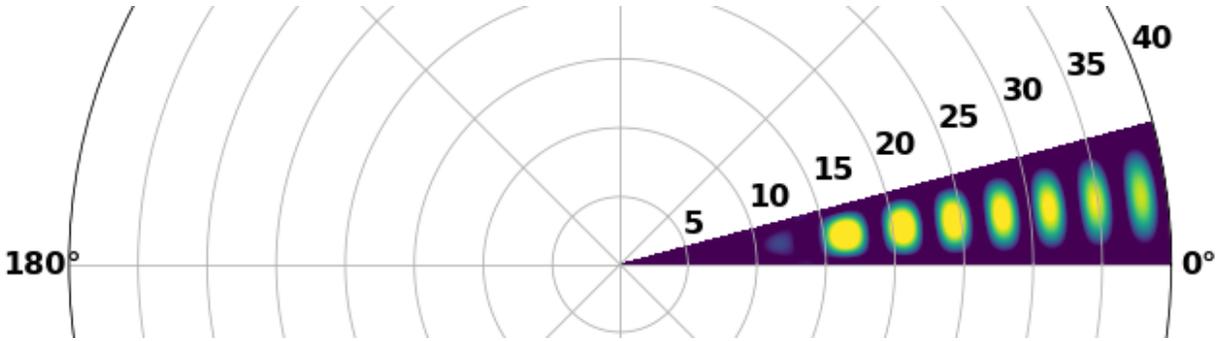

a)

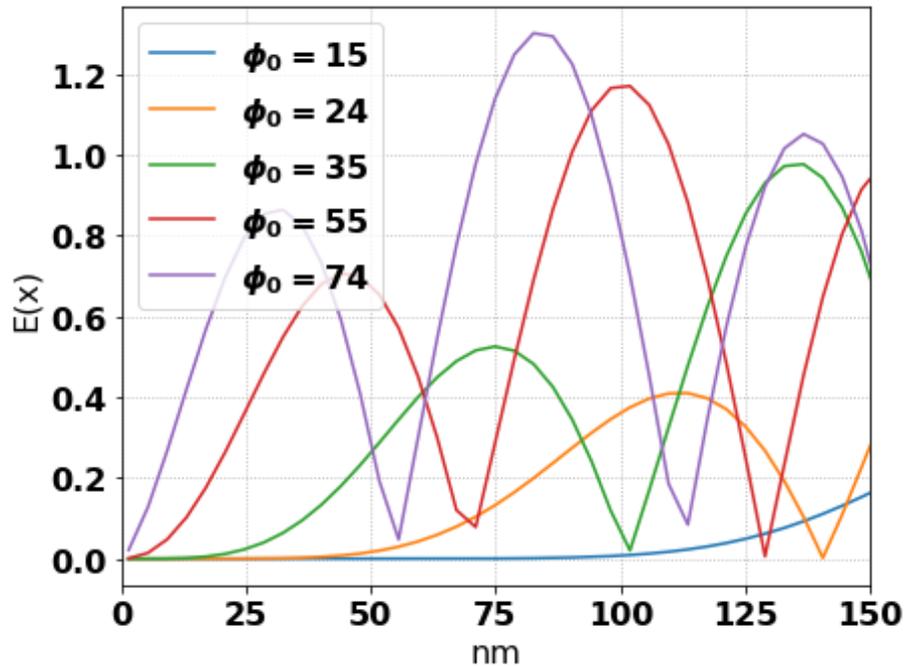

b)

**Supplementary Data Figure SD4) Study of profile cut of calculated localized optical hotspot by solving Maxwell's equations.** Localized hotspot of EF **(a)** and profile cut of calculated EF for v-shape structure with variety of corner angles **(b)** from Maxwell's equation solution in SD S.2 showing suppression of localized EF in the V-shape gap with 35⁰ around 18 nm away from the V-shape vertex (green color profile). This suppression is at area of V-shape structure with 11 nm opening width (in comparison to 13 nm width experimental observation of suppression of EF). The walls in V-shape structure for this theoretical calculation are considered as hard walls.

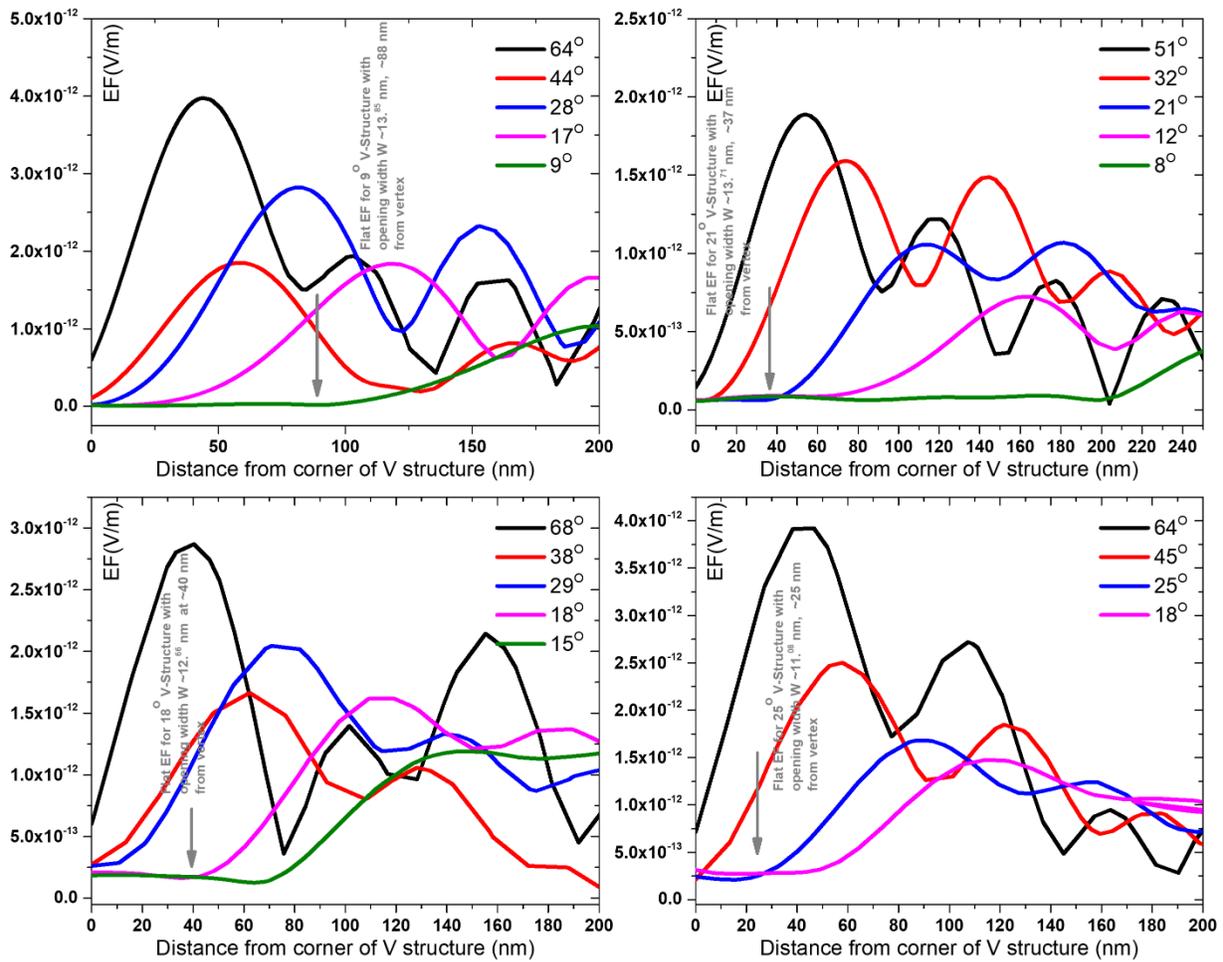

**Supplementary Data Figure SD5) Study of the suppression of light (Electric Field) by simulating localized light in V-shape structure and estimating lowest spatial limit of localization of light.** Profile cut of simulated EF for variety V-shape structure (presenting scattering dislocation structures) at different orientation, configuration and distance from incoming light. The localized EF is suppressed when the V-shape with opening gap width approaches ~11-13 nm. The localized hotspots are all plotted at 390 nm (like Fig.3). EF stays suppressed after the angle of V-shape structure becomes smaller than this critical gap.

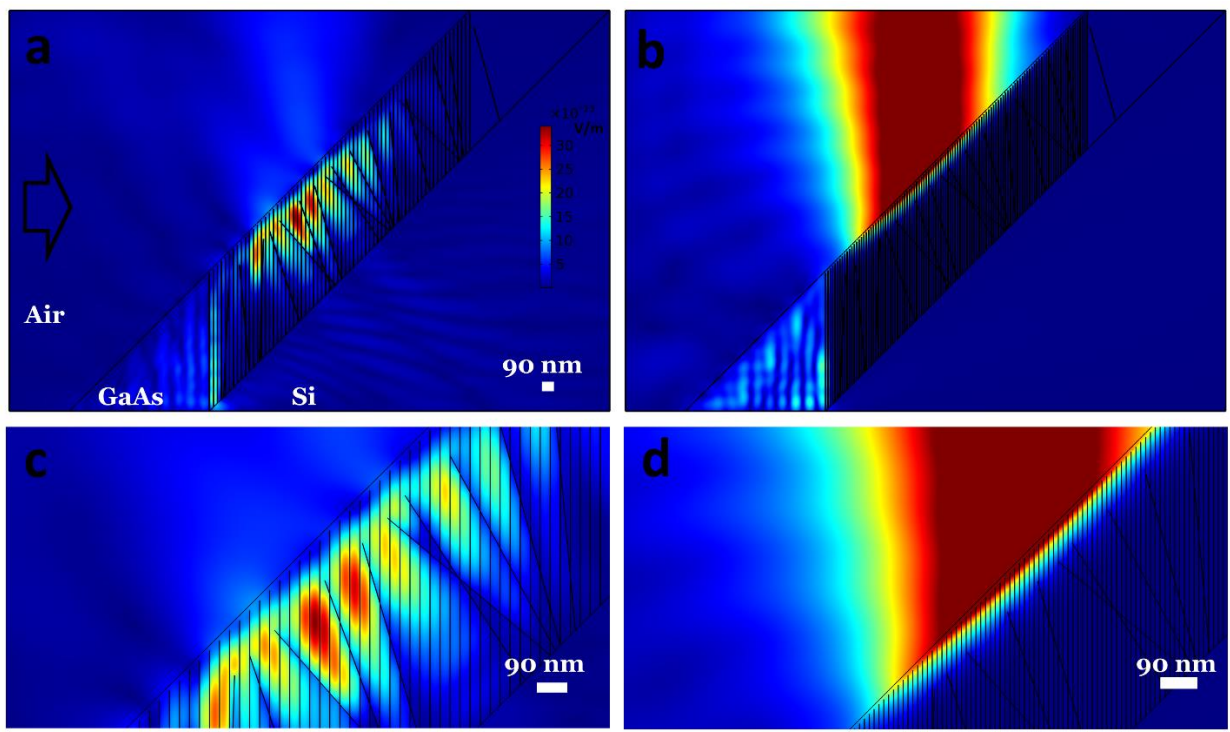

**Supplementary Data Figure SD6) Simulation of SHG localization of light in disordered medium with ~20 and ~12 nm dislocation gap (optical mean free path). (a)** Fundamental light (780 nm) penetrates into the scattering medium (GaAs with dislocations) while the scattering dislocation has a ~20 nm gap (with low randomness) and gets localized. Plotted SHG nonlinear response of the film (390 nm) hotspots shows localization of light in disordered scattering areas. **(b)** Localization of light gets suppressed in medium with ~12 nm gap dislocations distribution (with low randomness) and light gets scattered off of the sample as light can not penetrate in the film. The left lower corner of the GaAs film in (b) does show localization of light in a wide V-shape corner of the medium. **(c,d)** The lower plots show details of both plots of localization and suppression of light for a random medium with average 20 and 12 nm gap between scattering dislocations. Plots (b) and (d) replicate the $In_{30}Ga_{70}As$-GaAs film with 13 nm scattering gaps in Fig. 1 f and g). The arrow in (a) shows direction of propagation of fundamental excitation light.

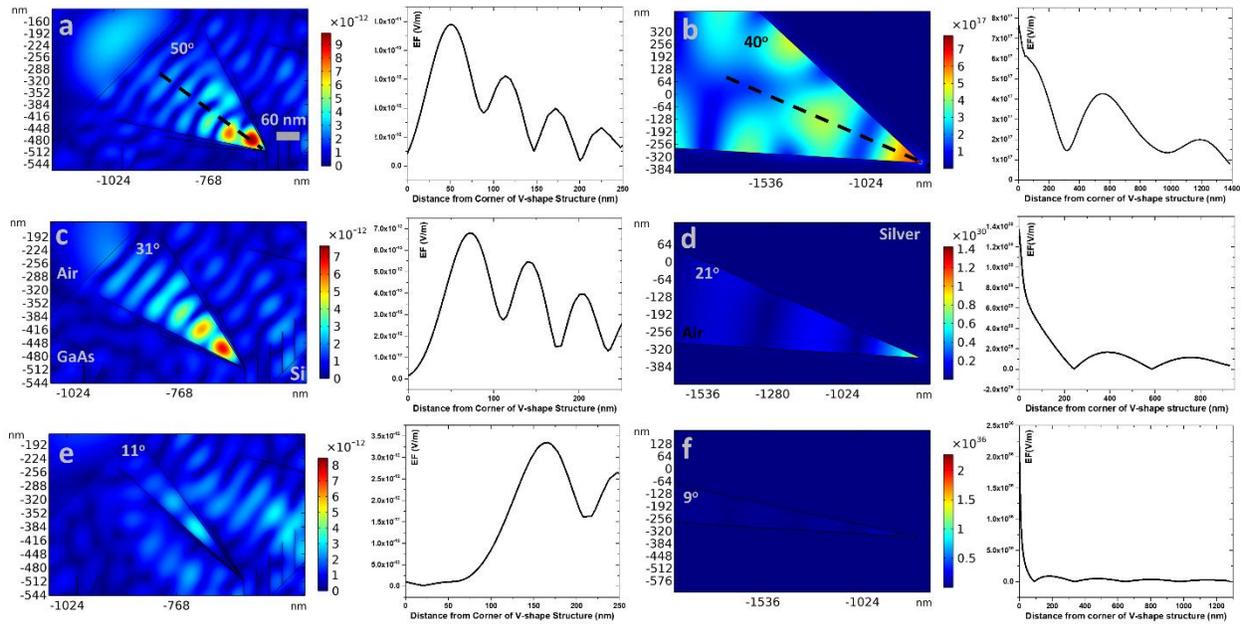

**Supplementary Data Figure SD7) Comparison of Anderson localization of light to plasmonic confinement and enhancement of light.** The localized light/EF simulation in scattering V-shape structure dislocation sites show that the fields get suppressed at corner of structure while the angle decreases from 50° to 11° **(a,c,e)** The electric-field (EF) profiles cuts are plotted from corner of structure vertex. The excitation light is at 780 nm and the plots are EF at SHG response of the structure at 390 nm. In comparison a metallic V-shape structure with air gap between the structure arm is studied in a variety of angles (40°, 21° and 9°) **(b,d,f)**. Plasmonic enhancement of light/EF at 780 nm is plotted inside the V-shape air gap. The EF simulation and the profile cuts show strong enhancement at vertex of the plasmonic structure, opposite to what was observed for localized light at vertex of scattering structure. In Metallic-Air V-shape structure, decreasing the vertex angle causes the EF enhancement at vertex area by orders of magnitude.